\documentclass{article}

\usepackage{Sweave}

\usepackage{geometry}
\usepackage[figuresright]{rotating}
\usepackage{tabu}
\usepackage{amsmath}
\usepackage{amssymb}
\usepackage{amsthm}
\usepackage{mathtools}
\usepackage{pifont}
\usepackage{lineno}
\usepackage{graphicx}
\usepackage{enumerate}
\usepackage{rotating}
\usepackage[round]{natbib}
\usepackage{scrextend}
\usepackage{datetime}
\usepackage{array}
\usepackage{multicol}
\usepackage[justification = raggedright]{subcaption}
\newcommand{\proglang}[1]{{\textsf{#1}}}
\newcommand{\R}{\proglang{R }}
\newcommand{\pkg}[1]{{\normalfont\fontseries{b}\selectfont #1}}
\newcommand{\pbs}[1]{\let\tmp\\#1\let\\\tmp} 
\newcommand{\s}{\mathbf{s}}
\newcommand{\h}{\mathbf{h}}

\newcommand{\A}{\mathbf{A}}
\newcommand{\G}{\mathbf{G}}
\newcommand{\D}{\mathcal{D}}

\newcommand{\bg}{\boldsymbol{\gamma}}

\newcommand{\bL}{\mathbf{\Lambda}}

\newtheorem{definition}[subsubsection]{Definition}
\numberwithin{equation}{section}

\title{spTest: An R Package Implementing Nonparametric Tests of Isotropy}
\author{Zachary D. Weller}

\begin{document}

\maketitle

\begin{abstract}
An important step of modeling spatially-referenced data is appropriately specifying the second order properties of the random field. A scientist developing a model for spatial data has a number of options regarding the nature of the dependence between observations. One of these options is deciding whether or not the dependence between observations depends on direction, or, in other words, whether or not the spatial covariance function is isotropic. Isotropy implies that spatial dependence is a function of only the distance and not the direction of the spatial separation between sampling locations. A researcher may use graphical techniques, such as directional sample semivariograms, to determine whether an assumption of isotropy holds. These graphical diagnostics can be difficult to assess, subject to personal interpretation, and potentially misleading as they typically do not include a measure of uncertainty. In order to escape these issues, a hypothesis test of the assumption of isotropy may be more desirable. To avoid specification of the covariance function, a number of nonparametric tests of isotropy have been developed using both the spatial and spectral representations of random fields. Several of these nonparametric tests are implemented in the \R package \pkg{spTest}, available on \texttt{CRAN}. We demonstrate how graphical techniques and the hypothesis tests programmed in \pkg{spTest} can be used in practice to assess isotropy properties.
\end{abstract}

\section{Introduction}
An important step in modeling a spatial process is choosing the form of the covariance function. The form of the covariance function will have an effect on kriging as well as parameter estimates and the associated uncertainty \citep[pg.~127-135]{cressie1993}. A common simplifying assumption about the spatial covariance function is that it is isotropic, that is, the dependence between sampling locations depends only on the distance between locations and not on their relative orientation. This assumption may not always be reasonable; for example, wind may lead to directional dependence in environmental monitoring data. Misspecification of the second order properties can lead to misleading inference. \citet[pg.~87-90]{sherman2011spatial} and \citet{guan2004isotropy} summarize the effects of incorrectly specifying isotropy properties on kriging estimates through numerical examples. In order to choose an appropriate model, a statistician must first assess the nature of the spatial variation of his or her data. To check for anisotropy (directional dependence) in spatially-referenced data, a number of graphical techniques are available, such as directional sample variograms or rose diagrams \citep[pg.~149-154]{matheron1961precision,isaaks1989applied}. Spatial statisticians may have intuition about the interpretation and reliability of these diagnostics, but a less experienced user may desire evaluation of assumptions via a hypothesis test. In this paper, we use two real data examples to illustrate how the nonparametric hypothesis tests available in the \R \citep{rsoftware} package \pkg{spTest} \citep{spTest} can be used to assess isotropy properties in spatially-referenced data. The examples also demonstrate how graphical techniques and hypothesis tests can be used in a complementary role. The remainder of the paper is organized as follows: Section \ref{notation} establishes notation and definitions; Section \ref{sptest-functions} describes the functionality of the \pkg{spTest} package; Section \ref{applications} demonstrates how to use the functions in \pkg{spTest} in conjunction with graphical techniques on two different data sets; Section \ref{discussion} concludes the paper with a discussion.

\section{Notation}\label{notation}

\subsection{Spatial statistics}\label{spatial}
In Section \ref{spatial} we briefly review key definitions required for describing and performing tests of isotropy. For additional background, see \citet{weller2015} or \citet{schab}. Let $\{Y(\s): \s \in \D \subseteq \Re^2\}$ be a second order stationary random field (RF). Let $\{\s_1, \ldots, \s_n\} \subset \D$ be the finite set of locations at which the random process is observed, providing the random vector $\mathbf{Y} = (Y(\s_1), \ldots , Y(\s_n))^\top$. The sampling locations may follow one of several spatial sampling designs, for example, gridded locations, randomly and uniformly distributed locations, a cluster design, or any other general design. Note that there is a distinction between a lattice process and a geostatistical process observed on a grid \citep[e.g.,][pg.~6-10]{fuentes2010spectral, schab}, but we do not explore this distinction and will use the term grid throughout.

For a spatial lag $\h = (h_1, h_2)^\top$, the semivariogram function describes dependence between observations, $\mathbf{Y}$, at spatial locations separated by lag $\h$ and is defined as
\begin{equation}\label{semivariogram}
	\bg(\h) = \frac{1}{2}\mbox{Var}(Y(\s + \h) - Y(\s)).
\end{equation}
The classical moment-based estimator of the semivariogram \citep{matheron1962traite} is 
	\begin{equation}\label{classical}
		\widehat{\gamma}(\h) = \frac{1}{2|\D(\h)|} \sum [Y(\s) - Y(\s + \h)]^2,
	\end{equation}
where the sum is over $\D(\h) = \{\s:\s, \s + \h \in \D\}$ and $|\D(\h)|$ is the number of elements in $\D(\h)$. The set $\D(\h)$ is the set of sampling location pairs that are separated by spatial lag $\h$. The covariance function is defined by
\begin{equation}\label{covfcn}
 C(\h) = \mbox{Cov}(Y(\s), Y(\s + \h))
 \end{equation}
 and is an alternative to the semivariogram for describing spatial dependence. When it exists, the covariance function, $C(\h)$, is related to the semivariogram by $C(\h) = lim_{||\mathbf{v}|| \rightarrow \infty}\bg(\mathbf{v}) - \bg(\h)$ if the limit exists.
 
It is often of interest to infer the effect of covariates on the process, deduce dependence structure, and/or predict $Y$ with associated uncertainty at new locations. To achieve these goals, the practitioner must specify the distributional properties of the spatial process. A common assumption is that the finite-dimensional joint distribution is multivariate normal (MVN), in which case we call the RF a Gaussian random field (GRF).

A common simplifying assumption about the spatial dependence structure is that it is isotropic.
\begin{definition}\label{iso}
	A second-order stationary spatial process is isotropic if the semivariogram, $\bg(\h)$, [or equivalently, the covariance function $C(\h)$] of the spatial process depends on the lag vector $\h$ only through its Euclidean length, $||\h||$, i.e., $\bg(\h) = \bg_{0}(||\h||)$ for some function $\bg_{0}(\cdot)$ of a univariate argument.
\end{definition}
\noindent Isotropy implies that the dependence between any two observations depends only on the Euclidean distance between their sampling locations and not on their relative orientation. A spatial process that is not isotropic is called anisotropic. The methods in \pkg{spTest} are designed to assist in determining whether or not the assumption of isotropy holds. Namely, the functions in \pkg{spTest} implement nonparametric hypothesis tests of isotropy, which avoid the choice of a parametric form for the covariance (semivariogram) function.

There are two important modifications to the estimator in Equation~\ref{classical} that are pertinent to the methods described in this paper. First, for non-gridded sampling locations, the estimator needs to be modified to account for the fact that very few or no pairs of locations will be separated by a specific spatial lag, $\h$. One solution to this challenge is to specify a distance tolerance, $\epsilon$, such that lags having length $||\h|| \pm \epsilon$ are included in estimating the semivariance at lag $\h$. Second, directional sample semivariograms can be estimated by only using observations that are separated by spatial lags in a specific direction. For example, to investigate potential anisotropy, we can compare sample semivariograms between the horizontal and vertical directions. For non-gridded sampling locations, very few pairs of locations will lie at a specific distance and directional lag, so we need to allow for both a distance and a directional tolerance when estimating the semivariance. A common method for doing this is by using a product kernel smoother that smoothes over both the horizontal ($h_1$) and vertical ($h_2$) components of the spatial lag $\h = (h_1, h_2)^\top$.

Spatial RFs and their second order properties can also be expressed in the spectral (or frequency) domain using Fourier transforms. The spectral representation of RFs and their second order properties provides alternative methods for testing second order properties. For brevity we focus only on the methods in the spatial domain and refer the interested reader to \citet{weller2015} or \citet{fuentes2010spectral}. We note that that, in addition to the methods from the spatial domain, the spectral methods from \citet{lu2005test} are also implemented in \pkg{spTest}.

\subsection{Nonparametric tests of isotropy}\label{nptests}
\citet{lu1994distributions} and \citet{lu2001testing} pioneered a popular approach to testing second-order properties when they used the asymptotic joint normality of the sample semivariogram computed at different spatial lags to evaluate symmetry and isotropy properties. The subsequent works of \citet{guan2004isotropy, guan2007asymptotic} and \citet{maity2012test} built upon these ideas and are the primary methods programmed in \pkg{spTest}. Here we give an overview of the tests in \citet{guan2004isotropy} and \citet{maity2012test}.

Under the null hypothesis that the RF is isotropic, it follows that the values of $\bg(\cdot)$ evaluated at any two spatial lags that have the same norm are equal, independent of the direction of the lags. For example, under the assumption of isotropy, $\bg((-6,0)) = \bg\left(\left(\sqrt{3}, \sqrt{3} \hspace{1 mm} \right)\right)$. To completely specify the null hypothesis of isotropy, theoretically, one would need to compare semivariogram values for an infinite set of lags. In practice, a small number of lags are specified. Then it is possible to test a hypothesis consisting of a set of linear contrasts of the form 
\begin{equation}\label{linearhypothesis}
H_0: \A\bg(\cdot) = \mathbf{0}
\end{equation}
as a proxy for the null hypothesis of isotropy, where $\A$ is a full row rank matrix \citep{lu2001testing}. For example, a set of lags, denoted $\bL$, commonly used in practice on gridded sampling locations with unit spacing is
\begin{equation}\label{lagset}
 \bL = \{ \h_1 = (1,0), \h_2 = (0,1), \h_3 = (1,1),  \h_4 = (-1,1) \}, 
 \end{equation}
 and the corresponding $\A$ matrix under $H_0: \A\bg(\bL) = \mathbf{0}$ is
\begin{equation}\label{amat}
	\A = \begin{bmatrix}
		1 & -1 & 0 & 0 \\
		0 & 0 & 1 & -1 
		\end{bmatrix}.
\end{equation}

In other words, by using Equations~\ref{lagset} and \ref{amat} under the hypothesis \ref{linearhypothesis}, we contrast the semivariogram values at lags $\h_1 = (1,0)$ and $\h_2 = (0,1)$, and likewise, contrast semivariogram values at lags $\h_3 = (1,1)$ and $\h_4 = (-1,1)$. An important step in detecting anisotropy is the choice of lags, $\bL$, as the test results will only hold for the set of lags considered \citep{guan2004isotropy}. While this choice is somewhat subjective, there are several considerations for determining the set of lags. First, in terms of Euclidean distance, short lags should be chosen because estimates of the semivariance at long lags will be less reliable than estimates at short lags because they are based on fewer pairs of observations and hence more variable. Second, pairs of orthogonal lags should be contrasted because, for an anisotropic process, the directions of strongest and weakest spatial correlation will typically be orthogonal. For other considerations and more detailed guidelines regarding the choice of lags, see\citet{weller2015}, \citet{lu2001testing}, and \citet{guan2004isotropy}.

The tests in \citet{guan2004isotropy} and \citet{maity2012test} involve estimating either the semivariogram \ref{semivariogram} or covariogram \ref{covfcn} and evaluating the estimator at the set of chosen lags. Denoting the set of point estimates of the variogram/covariogram at the chosen lags as $\widehat{\G}_n$, the true values as $\G$, and normalizing constant $a_n$, a central result for all three methods is that
\begin{equation}\label{asymptoticMVN}
a_n(\widehat{\G}_n - \G) \xrightarrow[n \rightarrow \infty]{d} MVN(\mathbf{0}, \mathbf{\Sigma}),
\end{equation}
under increasing domain asymptotics and mild moment and mixing conditions on the RF. The test statistic is a quadratic form
\begin{equation}\label{ts}
TS = b_n^2(\A\widehat{\G}_n)^{\top}(\A\widehat{\mathbf{\Sigma}}\A^{\top})^{-1} (\A\widehat{\G}_n),
\end{equation}
where $\widehat{\mathbf{\Sigma}}$ is an estimate of the asymptotic variance-covariance matrix and $b_n$ is a normalizing constant. A $p$~value can be obtained from the asymptotic $\chi^2$ distribution with degrees of freedom given by the row rank of $\A$. Important differences in these works regard the distribution of sampling locations, shape of the sampling domain, estimation of $\G$ and $\mathbf{\Sigma}$, and assumptions on the dependence structure of the random field (see \citet{weller2015} for more details).

\section{Nonparametric tests implemented in spTest}\label{sptest-functions}
The \R package \pkg{spTest} includes functions for implementing the tests developed in \citet{guan2004isotropy}, \citet{lu2005test}, and \citet{maity2012test}. The primary functions in \pkg{spTest} for implementing these tests are \verb@LuTest@, \verb@MaityTest@, \verb@GuanTestGrid@, and \verb@GuanTestUnif@. For example, the test from \citet{guan2004isotropy} for data observed at non-gridded, but uniformly distributed, sampling locations is implemented in the function \verb@GuanTestUnif@, which takes the following arguments:

\begin{center}
\begin{verbatim}
GuanTestUnif(spdata, lagmat, A, df, h = 0.7, kernel = "norm",
  truncation = 1.5, xlims, ylims, grid.spacing = c(1, 1),
  window.dims = c(2, 2), subblock.h, sig.est.finite = T).
\end{verbatim}
\end{center}
There are several necessary inputs. The matrix \texttt{spdata} includes the coordinates of sampling locations and the corresponding data values. The spatial lags that will be used to estimate the semivariance, denoted $\bL$, are specified in the matrix \texttt{lagmat}. The matrix \texttt{A} provides the contrasts of the semivariance estimates, and its row rank is indicated by the parameter \texttt{df} (the degrees of freedom for the asymptotic $\chi^2$ distribution). The values \texttt{h} and \texttt{kernel} provide the bandwidth (smoothing) parameter and form of the kernel smoother, respectively, used to smooth over spatial lags when estimating the semivariance. If a normal smoothing kernel is used, then the \texttt{truncation} parameter indicates where to truncate the normal kernel (i.e., zero weight for values larger than this value). The parameters \texttt{xlims} and \texttt{ylims} give the horizontal and vertical limits of the sampling region (a rectangular sampling region is assumed). A grid is laid over the sampling region according the width and height specified by \texttt{grid.spacing}. The dimensions of the moving windows, given in the units of the underlying grid, are determined by the values in \texttt{window.dims}. The bandwidth used to estimate the semivariance on the subblocks of data is indicated by \texttt{subblock.h} and a finite sample adjustment to the estimate of the asymptotic variance-covariance matrix is made by setting \texttt{sig.est.finite = T}. For more information about the different arguments and guidelines on how to choose them, see \citet{weller2015}, the \pkg{spTest} manual, and the original works.

\section{Applications: Using spTest to check for anisotropy}\label{applications}
We demonstrate the functions in \pkg{spTest} on two data sets: the first containing gridded sampling locations, the second non-gridded sampling locations. For more details on the functions and more examples using simulated data, see the official \pkg{spTest} manual. The \pkg{spTest} package can be used independently of other packages built for analyzing spatial data, but it works nicely with two other packages loaded into \R: \pkg{fields} \citep{fields} and \pkg{geoR} \citep{geor}. The package \pkg{fields} is automatically installed with \pkg{spTest}, while \pkg{geoR} is not. We also load the \pkg{splines} package \citep{splines}, which we use to fit mean functions.

\begin{Schunk}
\begin{Sinput}
R> library("spTest")
R> library("fields")
R> library("geoR")
R> library("splines")
\end{Sinput}
\end{Schunk}

For the two examples given below, we use graphical diagnostics and the hypothesis tests implemented in \pkg{spTest} to determine whether or not an assumption of isotropy is reasonable for spatially-referenced data. The general strategy will be to first do exploratory data analysis (EDA) of the original data and create a model for the mean of the spatial process using appropriate covariates. After estimating a model for the mean, we extract residuals and again use EDA to check for remaining spatial dependence and utilize graphical diagnostics and hypothesis tests to investigate potential anisotropy. For brevity, we have not included the full version of EDA code and plots; instead, we include only the most relevant to demonstrating the functionality of the \pkg{spTest} package. The complete version of the code is available on \texttt{github}.

\subsection{Gridded sampling locations}
\begin{figure}
	\begin{subfigure}[t]{0.5\linewidth}
	\centering
	\includegraphics[width = 2.5 in]{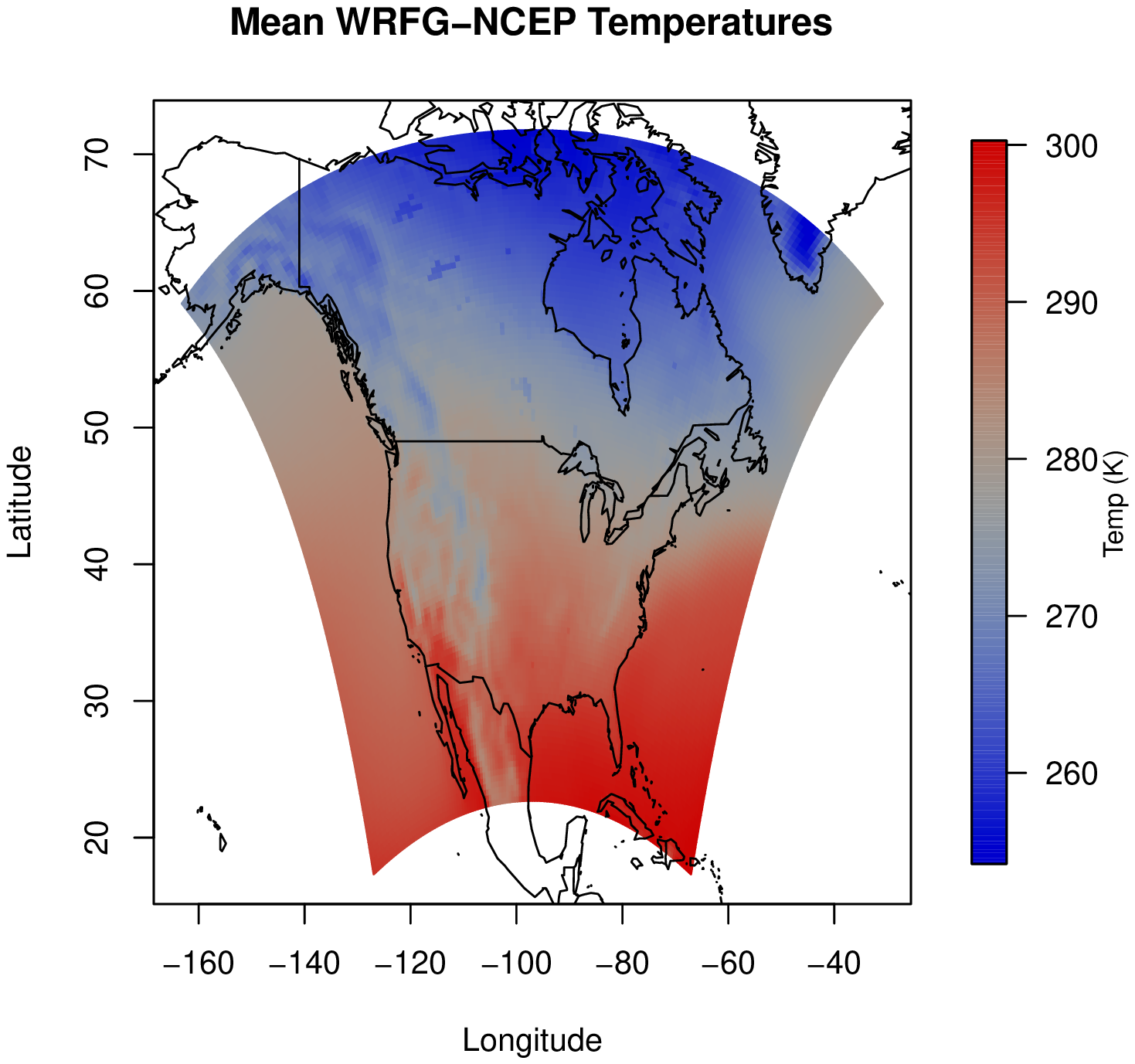}
	\caption{}
	\label{narccap1}
	\end{subfigure}
	\begin{subfigure}[t]{0.5\linewidth}
	\centering
	\includegraphics[width = 2.5 in]{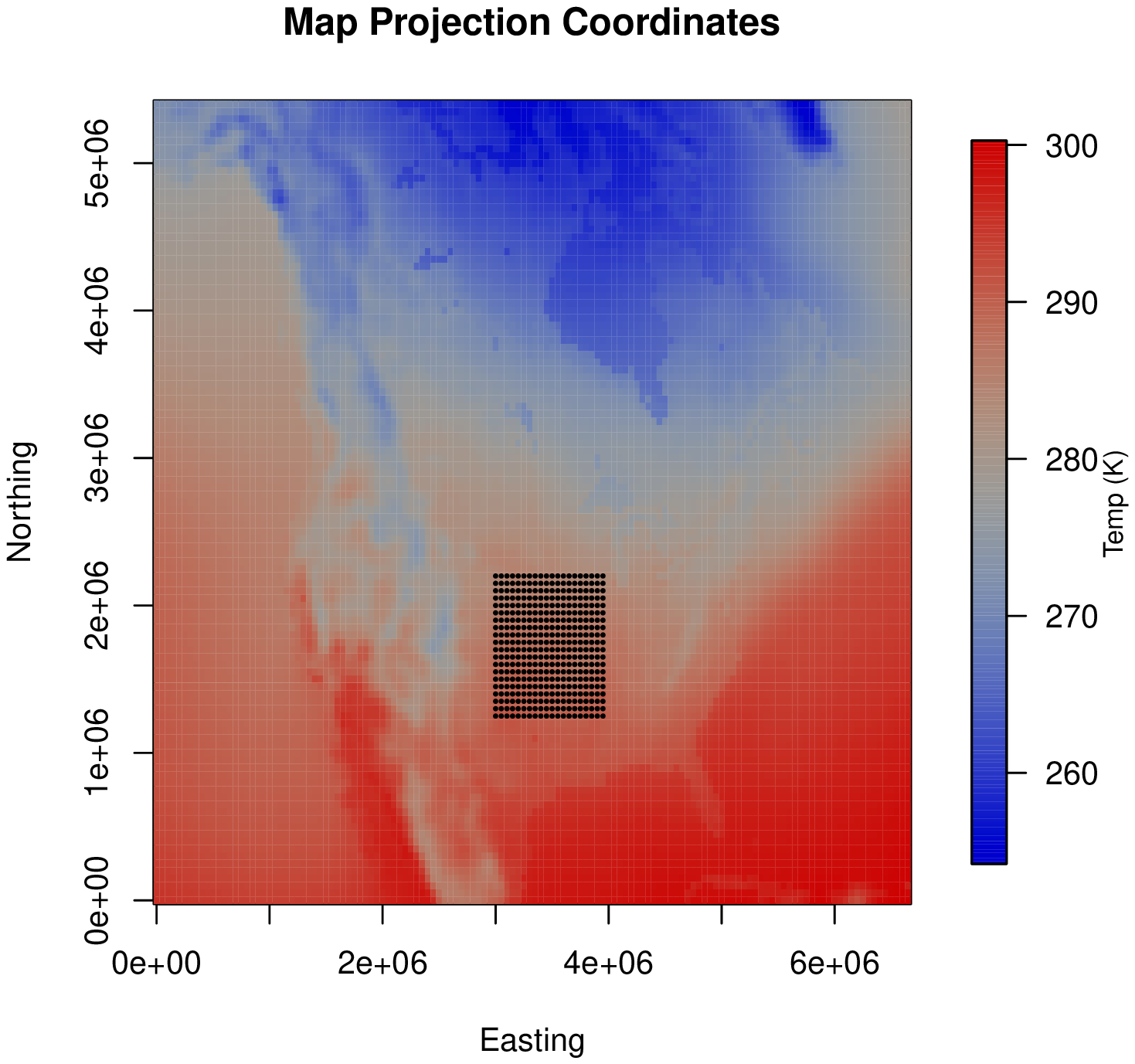}
	\caption{}
	\label{narccap2}
	\end{subfigure}
\caption{Heat maps of average temperatures of WRFG-NCEP output from NARCCAP. Figure~\ref{narccap1} displays the average of WRFG-NCEP yearly average temperature over 24 years with coordinates in latitude and longitude. Figure~\ref{narccap2} displays the same data but on the gridded map projection (easting/northing) coordinates, and the black points indicate the subset of the data used for this example. }
\label{narccap}
\end{figure}

The gridded data used in this section come from the North American Regional Climate Change Assessment Program [NARCCAP] \citep{mearns2009regional}. The data set \texttt{WRFG} in \pkg{spTest} includes coordinates and a 24-year average of yearly average temperatures from runs of the Weather Research and Forecasting - Grell configuration (WRFG) regional climate model (RCM) using boundary conditions from the National Centers for Environmental Prediction (NCEP). The original data and the \R code used to create the yearly averages are available online. The data set contains both latitude and longitude coordinates and map projection coordinates that specify the regular grid for 14,606 grid boxes along with average temperature at each grid box. We can display a heat map of all of the data using the \texttt{image.plot} function from the \pkg{fields} package (see Figure~\ref{narccap1}). Due to computational considerations and because the methods in \pkg{spTest} assume stationarity, for our analysis we use a $20 \times 20$ subset of the grid boxes over the central United States (see Figure~\ref{narccap2}).

\begin{Schunk}
\begin{Sinput}
R> data("WRFG")
R> image.plot(WRFG$lon - 360, WRFG$lat, WRFG$WRFG.NCEP.tas, 
+      col = two.colors(n = 256, start = "blue3", end = "red3", 
+          middle = "gray60"), legend.lab = "Temp (K)", 
+      legend.cex = 0.8, legend.line = 2.2, xlab = "Longitude", 
+      ylab = "Latitude", main = "Mean WRFG-NCEP Temperatures")
R> world(add = T)
\end{Sinput}
\end{Schunk}
\begin{Schunk}
\begin{Sinput}
R> coords <- expand.grid(WRFG$xc,WRFG$yc)
R> sub <- which(coords[,1] > 2.95e6 & coords[,1] < 4.0e6 
+  & coords[,2] > 1.2e6 & coords[,2] < 2.25e6)
R> image.plot(WRFG$xc, WRFG$yc, WRFG$WRFG.NCEP.tas, 
+  col = two.colors(n = 256, start = "blue3", end = "red3", 
+  middle = "gray60"), legend.lab = "Temp (K)", legend.cex = 0.8, 
+  legend.line = 2.2, xlab = "Easting", ylab = "Northing", 
+  main = "Map Projection Coordinates")
R> points(coords[sub,], pch = 16, cex = 0.35)
R> my.coords <- coords[sub,]
R> colnames(my.coords) <- c("xc", "yc") 
R> tas <- WRFG$WRFG.NCEP.tas
R> tas <- c(tas)
R> tas <- tas[sub]
R> mydata <- cbind(my.coords[,1],my.coords[,2], tas)
\end{Sinput}
\end{Schunk}

To investigate potential anisotropy in the relevant subset of these data, we can examine two graphical diagnostics: a heat map and directional sample semivariograms. We use the function \texttt{variog4} from the \pkg{geoR} package to estimate directional semivariograms to visually assess isotropy properties.

\begin{Schunk}
\begin{Sinput}
R> x.coord <- unique(my.coords[, 1])
R> y.coord <- unique(my.coords[, 2])
R> nx <- length(x.coord)
R> ny <- length(y.coord)
R> tas.mat <- matrix(mydata[, 3], nrow = nx, ncol = ny, 
+      byrow = F)
R> image.plot(x.coord, y.coord, tas.mat, col = two.colors(n = 256, 
+      start = "blue3", end = "red3", middle = "gray60"), 
+      legend.lab = "Temp (K)", legend.cex = 0.8, legend.line = 2.2, 
+      ylab = "Northing", xlab = "Easting", main = "Subset of Temperatures")
\end{Sinput}
\end{Schunk}
\begin{Schunk}
\begin{Sinput}
R> geodat <- as.geodata(mydata)
R> svar4 <- variog4(geodat)
R> plot(svar4)
R> title("Directional Sample Semivariograms")
\end{Sinput}
\end{Schunk}
\begin{figure}
	\begin{subfigure}[t]{0.5\linewidth}
	\centering
	\includegraphics[width = 2.5 in, height = 2.5 in]{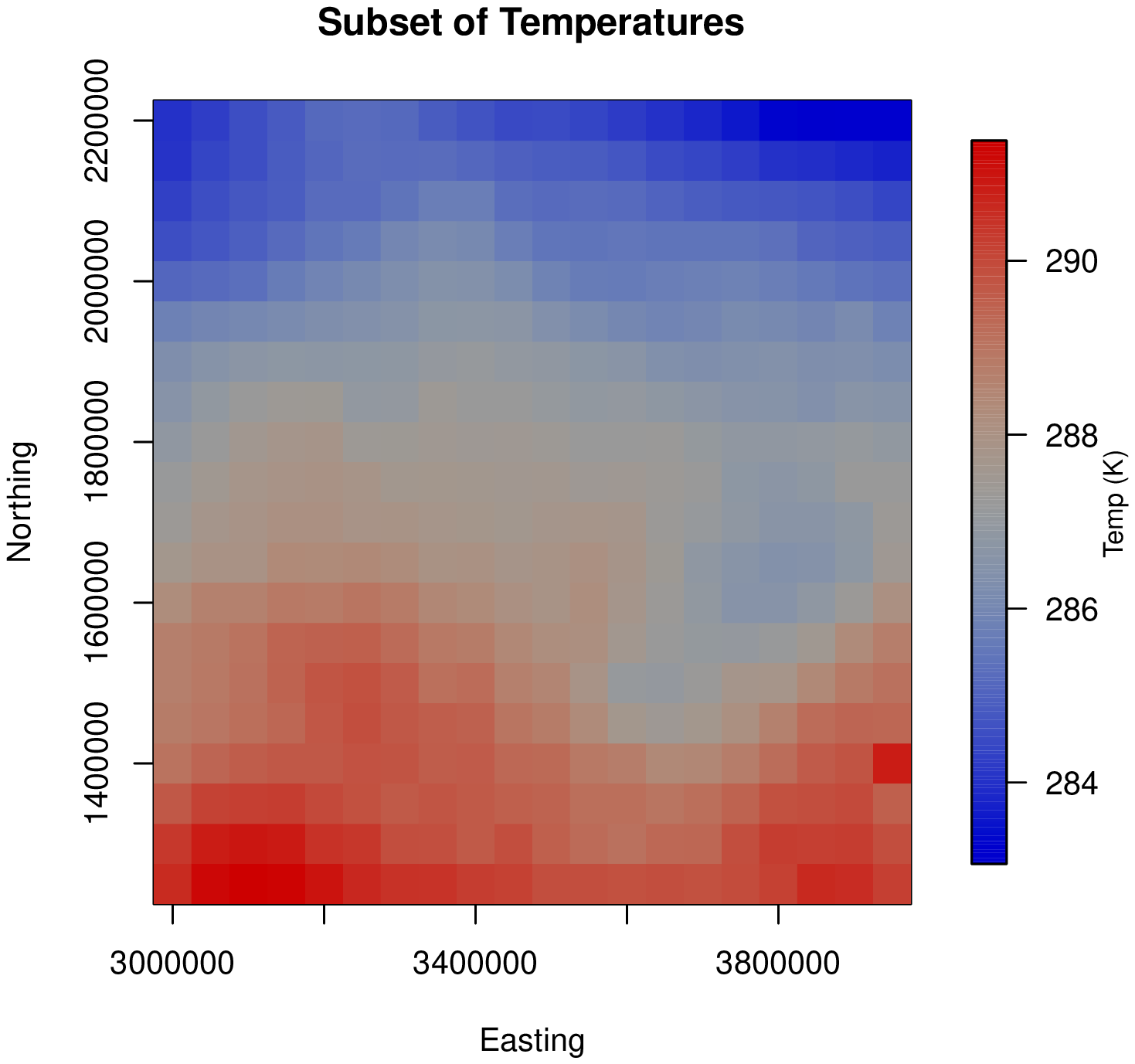}
	\caption{}
	\label{hm1}
	\end{subfigure}
	\begin{subfigure}[t]{0.5\linewidth}
	\includegraphics[width = 2.5 in]{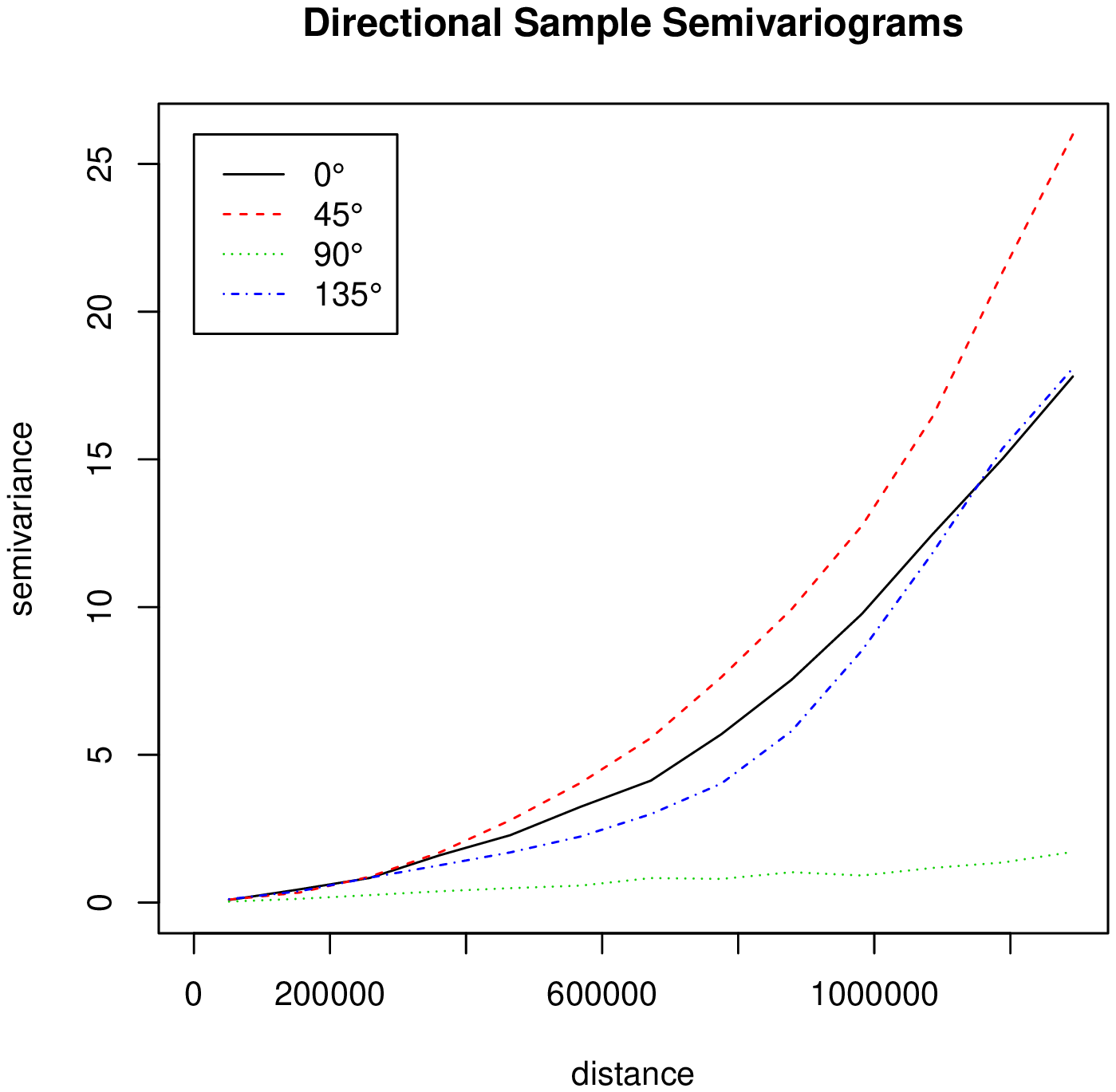}
	\caption{}
	\label{dv1}
	\end{subfigure}
\caption{Graphical assessments of isotropy in the $20 \times 20$ subset of WRFG temperature data. Because northing (latitude) coordinates have not been accounted for, the heat map (\ref{hm1}) indicates that the dependence between observations is stronger in the east-west direction than the north-south direction. The directional dependence is also evidenced by the differences between the directional sample semivariograms (\ref{dv1}).}
\label{fig2}
\end{figure}

The heat map in Figure~\ref{hm1} indicates that the spatial process is anisotropic, having a stronger spatial dependence in the horizontal direction than the vertical direction. Intuitively, northing (latitude) is an important factor in determining average temperature, and we need to include its effect in a model for these data. We also notice non-linear trends in temperature as a function of easting (longitude) in Figure~\ref{hm1}. Thus, the anisotropy can be attributed, at least in part, to the fact that we have not modeled important covariates related to the process. The directional sample semivariograms in Figure~\ref{dv1} reaffirm the notion that the data exhibit anisotropy as the $90^{\circ}$ sample semivariogram appears much different than the other three. Before modeling the effects of northing and easting coordinates, we use the \texttt{GuanTestGrid} function from \pkg{spTest} to affirm our understanding that these data exhibit anisotropy.

While some of the nonparametric methods for testing isotropy do not require Gaussian data, necessary conditions for the asymptotic properties of the test to hold are typically met when the data are Gaussian. A histogram (not shown) of the relevant subset of WRFG temperatures indicates that a Gaussian assumption is reasonable. To implement the nonparametric test in \citet{guan2004isotropy} via the function \texttt{GuanTestGrid}, we need to specify the spatial lags that will be used to test for differences in the semivariogram. For this test we choose the lag set 
\[
 \bL = \{ \h_1 = (1,0), \h_2 = (0,1), \h_3 = (1,1),  \h_4 = (-1,1)\},
\]
and we use the matrix $\A$ in Equation~\ref{amat} to contrast the semivariogram estimates. Because the distance between sampling locations is 50,000, we set the the scaling parameter \texttt{delta = 50,000}. To create subblocks of data used to estimate $\mathbf{\Sigma}$, we choose a moving window with size $4 \times 4$. The moving window dimensions should be chosen so that the window has the same shape (i.e., square or rectangle) and orientation as the sampling domain. To maximize the amount of data used to estimate $\mathbf{\Sigma}$, the dimensions of the window should evenly divide the number of columns and rows, respectively, of the entire region. The window dimensions should also be compatible with the spatial lags in $\bL$. For example, a window with dimensions of $2 \times 2$ cannot be used to estimate the variability of the semivariance at a lag with Euclidean distance longer than $\sqrt{2}$, the maximum distance between locations in the moving window. For this example there are 20 rows and columns and we are using lags with spacings of one or two sampling locations; hence, window dimensions of $2 \times 2$ or $4 \times 4$ are reasonable choices. We run the test via the following code.

\begin{Schunk}
\begin{Sinput}
R> my.delta <- 50000
R> mylags <- rbind(c(1, 0), c(0, 1), c(1, 1), c(-1, 1))
R> myA <- rbind(c(1, -1, 0, 0), c(0, 0, 1, -1))
R> tr <- GuanTestGrid(spdata = mydata, delta = my.delta, 
+      lagmat = mylags, A = myA, window.dims = c(4, 4), 
+      pt.est.edge = TRUE, sig.est.edge = TRUE, sig.est.finite = TRUE, 
+      df = 2)
R> tr$alternative <- NULL
R> tr
\end{Sinput}
\begin{Soutput}
	Test of isotropy from Guan et. al. (2004) for gridded
	sampling locations using the sample semivariogram.

data:  mydata
Chi-sq = 34.433, df = 2, p-value = 3.335e-08
p-value (finite adj.) < 2.2e-16, number of subblocks: 240

sample estimates: (lag value)
     (1,0)      (0,1)      (1,1)     (-1,1) 
0.03055723 0.08171415 0.10336776 0.10902089 

estimated asymp. variance-covariance matrix:
            [,1]        [,2]         [,3]        [,4]
[1,] 0.009229206 0.005124418  0.002365263  0.01657042
[2,] 0.005124418 0.032159967  0.016811371  0.04730438
[3,] 0.002365263 0.016811371  0.060613653 -0.01891585
[4,] 0.016570423 0.047304376 -0.018915852  0.12822978
\end{Soutput}
\end{Schunk}

As expected, the results of the hypothesis test ($p$~value $< 0.05$) indicate that the data exhibit anisotropy. We note that for gridded data, \citet{guan2004isotropy} recommend using the $p$~value computed via a finite sample correction. The function \texttt{GuanTestGrid}, and other functions in \pkg{spTest}, return a $p$~value(s) for the test and information used in computing the $p$~value, such as the point estimates ($\widehat{\G}_n$), estimates of the asymptotic variance-covariance matrix ($\widehat{\mathbf{\Sigma}}$), the number of subblocks used to estimate $\mathbf{\Sigma}$, and other information about the estimation process. Here we note that the point estimates for the directional semivariance are slightly different between the functions from the \pkg{spTest} and \pkg{geoR} packages due to different kernel methods used in estimation.

As mentioned earlier, we need to model the effects of northing and easting (latitude and longitude) coordinates on average temperature. We fit temperature as a nonparametric additive function of both the northing and easting coordinates via least-squares using cubic splines. The cubic splines can be specified using the function \texttt{ns} from the \pkg{splines} package and the least squares fit is computed via the \texttt{lm} function.

\begin{Schunk}
\begin{Sinput}
R> x1 <- my.coords[, 1]
R> x2 <- my.coords[, 2]
R> m1 <- lm(tas ~ ns(x1, df = 3) + ns(x2, df = 3))
R> summary(m1)
\end{Sinput}
\end{Schunk}

After removing the mean effects of the coordinates, we can check for any remaining (unaccounted for) spatial dependence and evidence of anisotropy in the residuals using graphical diagnostics and a hypothesis test. A histogram of the residuals (not shown) indicates that a Gaussian assumption is reasonable.

\begin{Schunk}
\begin{Sinput}
R> resid <- m1$resid
R> resid.mat <- matrix(resid, nrow = nx, ncol = ny, byrow = F)
R> image.plot(x.coord, y.coord, resid.mat, col = two.colors(n = 256, 
+      start = "blue3", end = "red3", middle = "gray60"), 
+      xlab = "Easting", ylab = "Northing")
R> title("Heat Map of Residuals")
\end{Sinput}
\end{Schunk}
\begin{Schunk}
\begin{Sinput}
R> resid.dat <- cbind(mydata[, 1:2], resid)
R> geodat.resid <- as.geodata(resid.dat)
R> plot(variog4(geodat.resid))
R> title("Directional Sample Semivariograms")
\end{Sinput}
\end{Schunk}
\begin{figure}
	\begin{subfigure}[t]{0.5\linewidth}
	\centering
	\includegraphics[width = 2.5 in, height = 2.5 in]{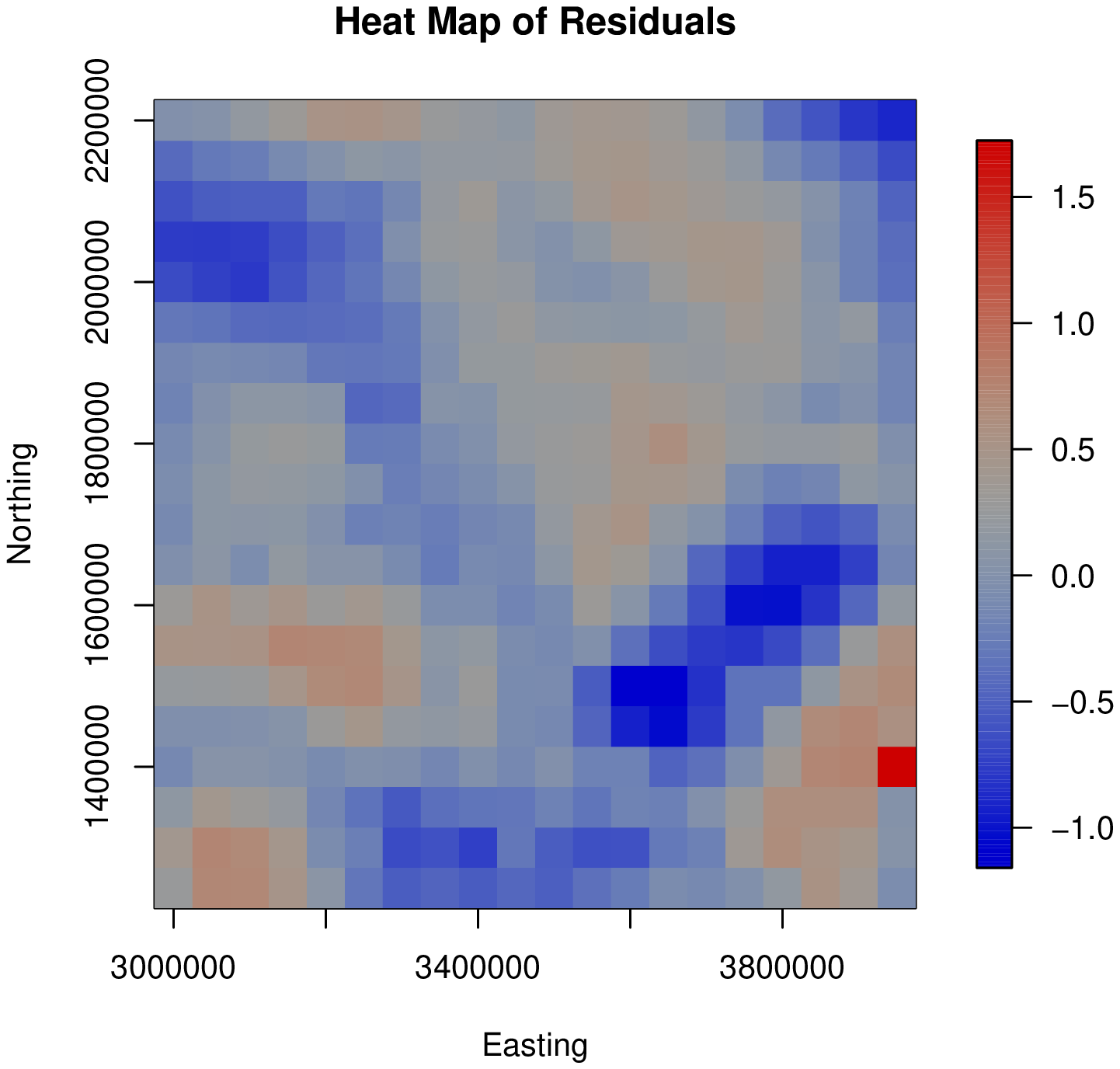}
	\caption{}
	\label{hm2}
	\end{subfigure}
	\begin{subfigure}[t]{0.5\linewidth}
	\includegraphics[width = 2.5 in]{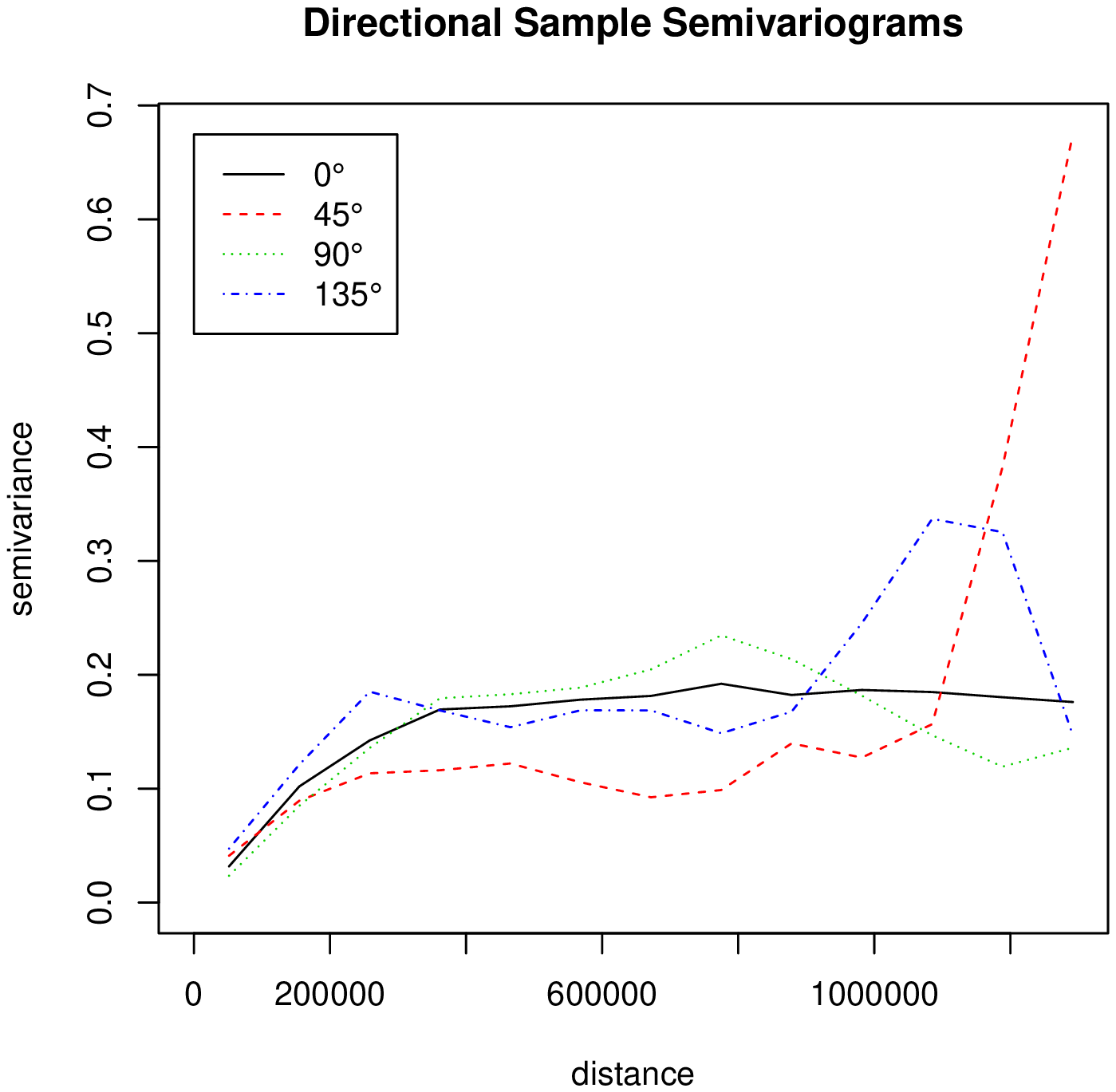}
	\caption{}
	\label{dv2}
	\end{subfigure}
\caption{Graphical assessments of isotropy in the residuals. The appearance of elongated areas of similar residual values in the heat map (\ref{hm2}) indicates that the process may be anisotropic. The directional semivariograms (\ref{dv2}) do not appear to exhibit differences, indicating that the process is isotropic.}
\label{fig3}
\end{figure}

The clusters of similar values in the heat map of Figure~\ref{hm2}, and the increase, followed by a leveling off, of the semivariance values as distance increases in the directional sample semivariograms in Figure~\ref{dv2} indicate that the residuals are still spatially dependent. However, the plots in Figure~\ref{fig3} do not clearly illustrate whether or not the residuals exhibit anisotropy. There appears to be directional dependence along the NW to SE direction in the northern parts of the heatmap (Figure~\ref{hm2}). The directional sample semivariograms do not appear to be different until the distance is greater than 200,000. Semivariogram estimates at large distances can be unreliable because there are fewer pairs of sampling locations to estimate this value than at short distances. Likewise, directional semivariograms are less reliable than a uni-directional semivariogram because fewer pairs of sampling locations are used at each distance for directional estimation. The unreliability of the semivariograms at the larger distances, coupled with the lack of a measure of uncertainty, make it difficult to determine whether or not an assumption of isotropy is reasonable. In hopes of gaining more insight into the isotropy properties, we perform a nonparametric hypothesis test of isotropy using the residuals with the same choices for $\bL$, $\A$, and the window dimensions.

\begin{Schunk}
\begin{Sinput}
R> tr <- GuanTestGrid(spdata = resid.dat, delta = my.delta, 
+      lagmat = mylags, A = myA, window.dims = c(2, 2), 
+      pt.est.edge = TRUE, sig.est.edge = TRUE, sig.est.finite = TRUE, 
+      df = 2)
R> tr$p.value.finite
\end{Sinput}
\begin{Soutput}
p.value.finite 
     0.4281046 
\end{Soutput}
\end{Schunk}

Here the residuals do not provide evidence for anisotropy ($p$~value $> 0.05$). These results suggest that it may be appropriate to choose an isotropic covariance function to model the residuals. However, it is important to note that we have not included the effect of other potentially influential covariates such as elevation or water cover in the model for temperature. Additionally, although we examined a $20 \times 20$ subset of the data, the grid boxes still cover a large geographic region of the U.S., and thus an assumption of stationarity, which is needed for the asymptotic properties of the hypothesis test to hold, may not be reasonable.

\subsection{Non-gridded sampling locations}
The non-gridded data set used in this section describes monthly surface meterology for the state of Colorado and comes from the National Center for Atmospheric Research (NCAR). The data are available in the package \pkg{fields}. For this example, our variable of interest is the log of the 30-year average of average yearly precipitation at 344 station locations during the time period 1968-1997. 

Like the temperature data, our goal will be to model the mean effect of covariates and check for spatial dependence and potential anisotropy in the residuals. Because the sampling locations cover a much smaller region than the subset of \texttt{WRFG} temperatures, we choose to use the latitude and longitude coordinates for this example. We can create a heat map of the log precipitation values and the elevation of the stations using the function \texttt{quilt.plot} from the \pkg{fields} package.

\begin{Schunk}
\begin{Sinput}
R> data("COmonthlyMet")
R> sub30 <- CO.ppt[74:103, , ]
R> nstations <- 376
R> years <- 1968:1997
R> nyears <- length(years)
R> yr.avg <- matrix(data = NA, nrow = nstations, ncol = nyears)
R> for (i in 1:nyears) {
+      yr.dat <- sub30[i, , ]
+      yr.avg[, i] <- apply(yr.dat, 2, mean, na.rm = T)
+  }
R> avg30 <- apply(yr.avg, 1, mean, na.rm = T)
R> quilt.plot(CO.loc, log(avg30), col = two.colors(n = 256, 
+      start = "blue3", end = "red3", middle = "gray60"), 
+      legend.lab = "Precip (log mm)", legend.cex = 0.8, 
+      legend.line = 2.2, xlab = "Longitude", ylab = "Latitude", 
+      main = "Quilt Plot of log(precip)")
R> US(add = T)
\end{Sinput}
\end{Schunk}
\begin{Schunk}
\begin{Sinput}
R> quilt.plot(CO.loc, CO.elev, col = two.colors(n = 256, 
+      start = "blue3", end = "red3", middle = "gray60"), 
+      legend.lab = "Elevation (meters)", legend.cex = 0.8, 
+      legend.line = 2.7, xlab = "Longitude", ylab = "Latitude", 
+      main = "Quilt Plot of Elevation")
R> US(add = T)
\end{Sinput}
\end{Schunk}

\begin{figure}
	\begin{subfigure}[t]{0.5\linewidth}
	\centering
	\includegraphics[width = 2.5 in]{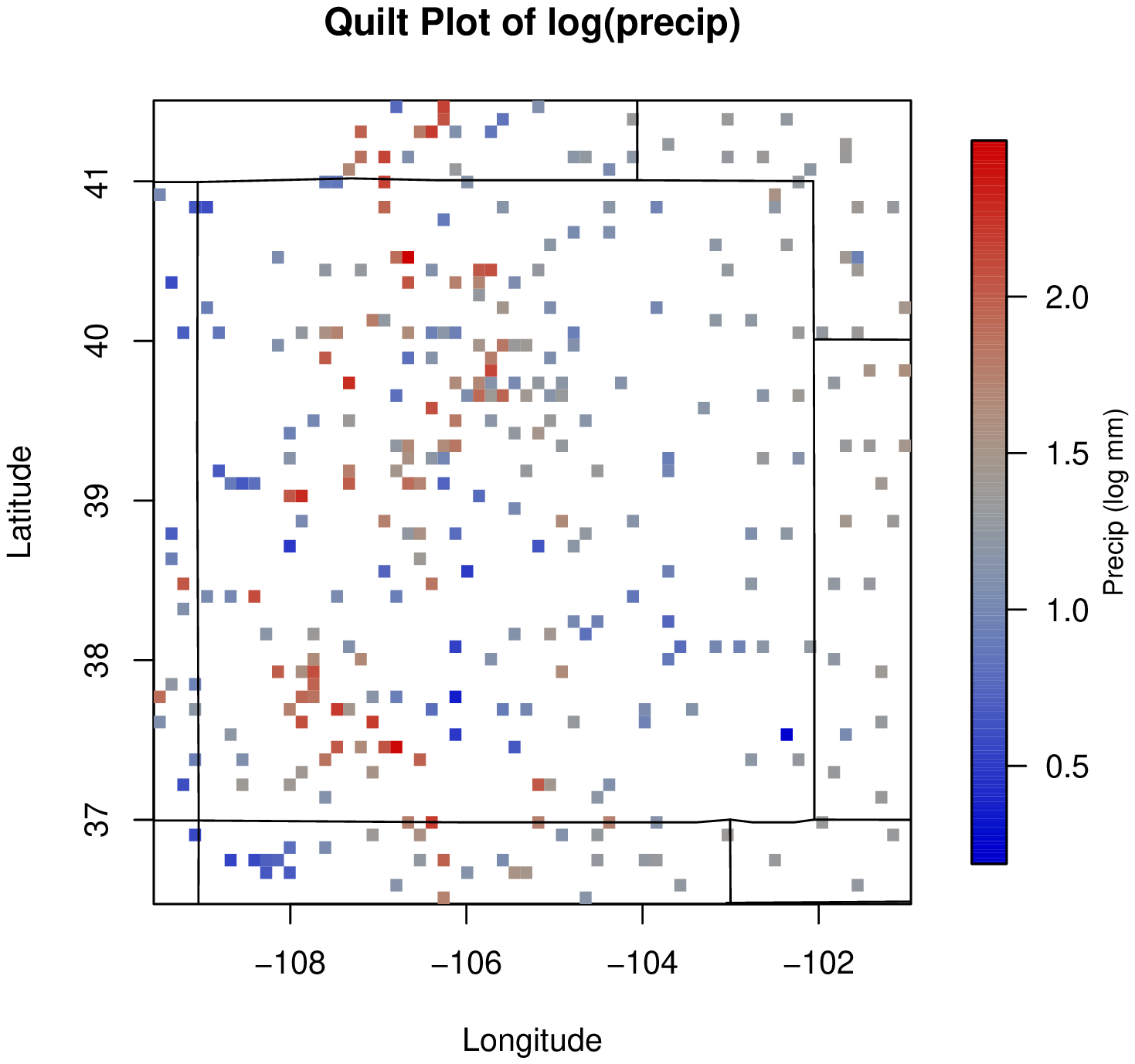}
	\caption{}
	\label{qp-precip}
	\end{subfigure}
	\begin{subfigure}[t]{0.5\linewidth}
	\includegraphics[width = 2.5 in]{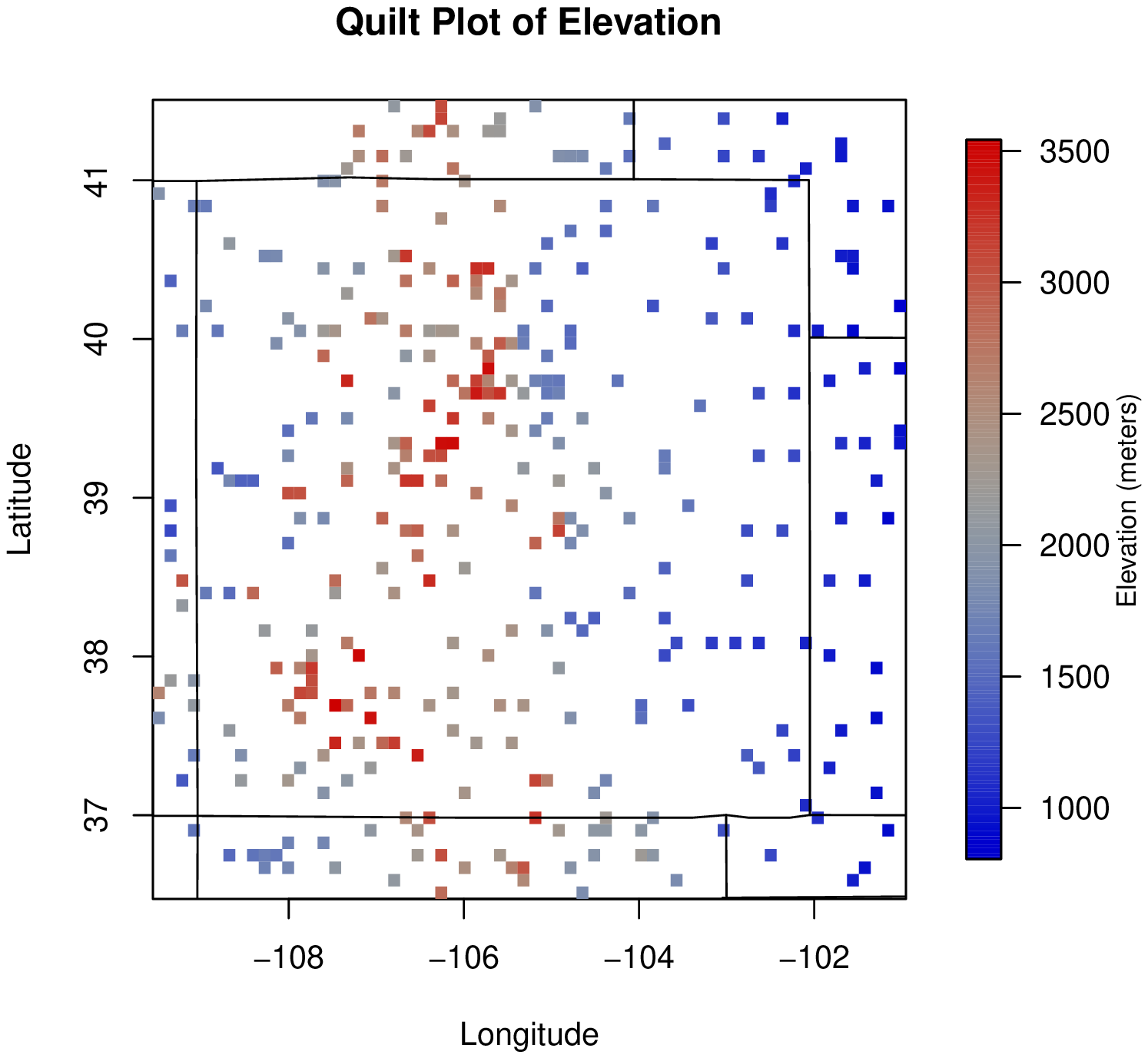}
	\caption{}
	\label{qp-elev}
	\end{subfigure}
\caption{Heat maps showing the locations of the stations in a region of Colorado along with the log of average precipitation (\ref{qp-precip}) and elevation (\ref{qp-elev}) at each station.}
\label{qplots}
\end{figure}

\begin{figure}
	\begin{subfigure}[t]{0.5\linewidth}
	\centering
	\includegraphics[width = 2.5 in]{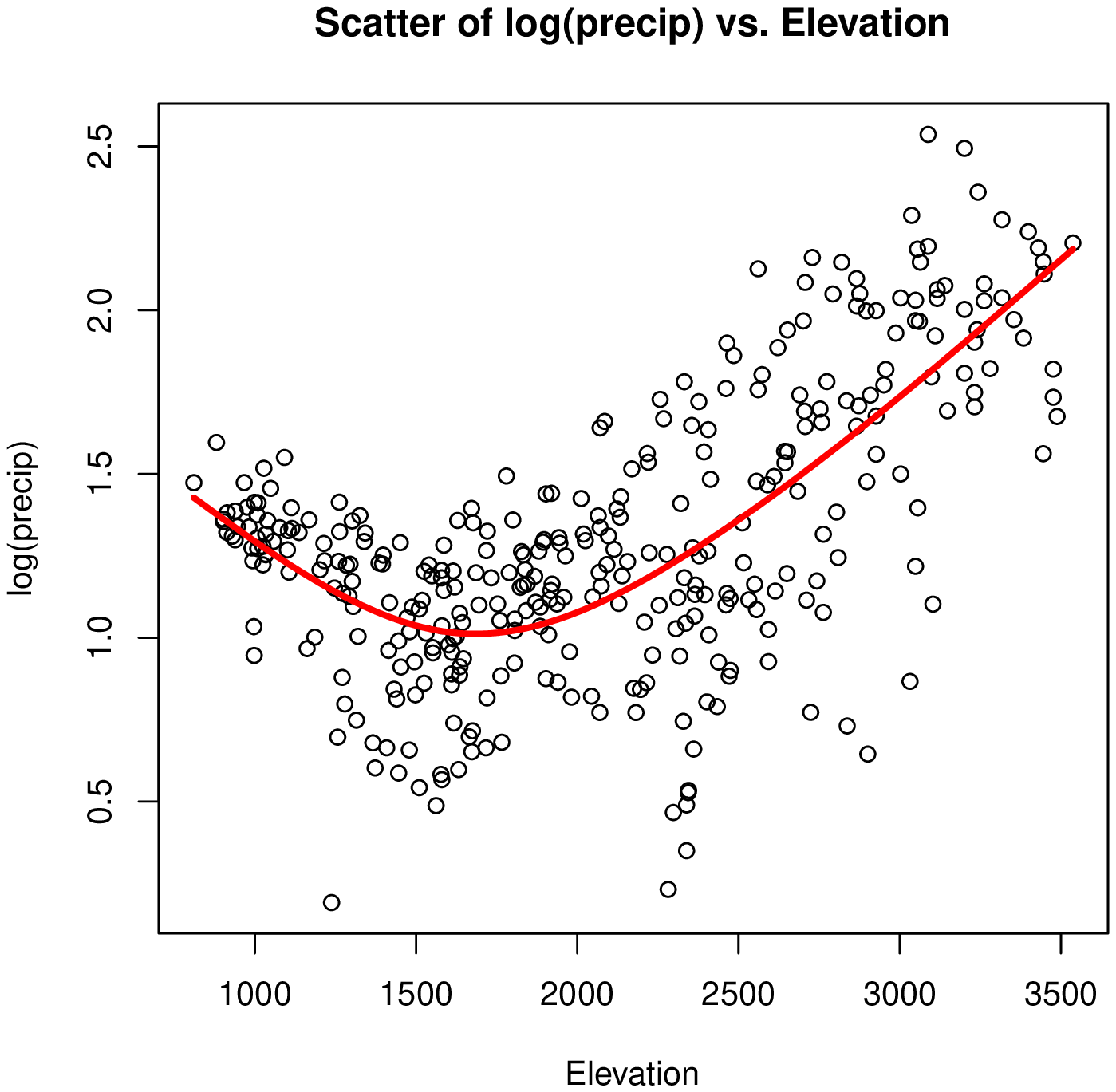}
	\caption{}
	\label{elev_scatter}
	\end{subfigure}
	\begin{subfigure}[t]{0.5\linewidth}
	\includegraphics[width = 2.5 in]{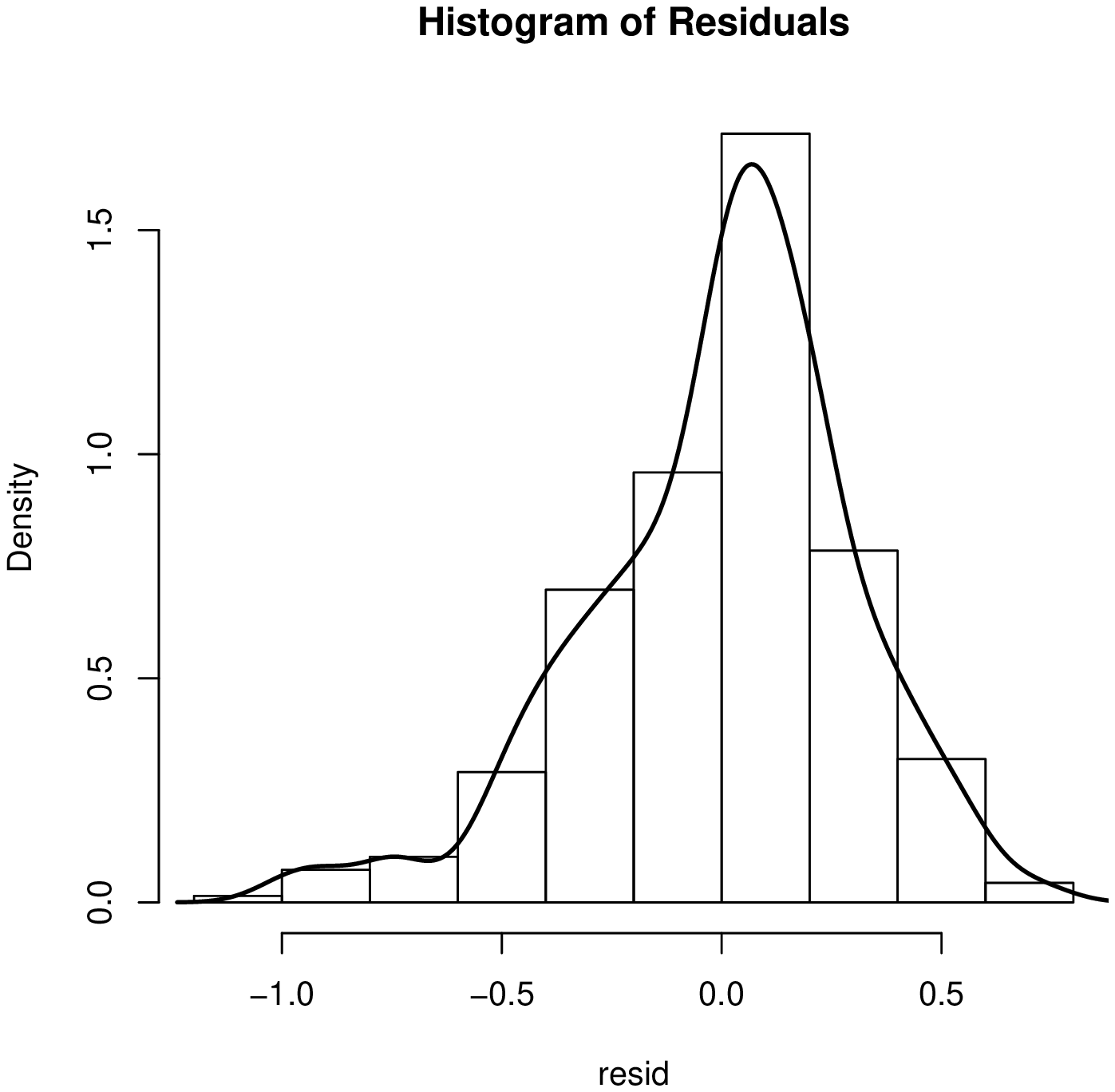}
	\caption{}
	\label{rhist}
	\end{subfigure}
\caption{Results from the model relating log(precipitation) and elevation.}
\label{rain_model}
\end{figure}

Colorado has two distinct geographic regions: the mountainous region in the west and the plains region in the east. Figure~\ref{qp-elev} illustrates these two regions, and we can begin to notice a possible relationship between elevation and average precipitation. We explore the potential relationship between precipitation and elevation using scatter plots (see Figure~\ref{elev_scatter}).

\begin{Schunk}
\begin{Sinput}
R> plot(CO.elev, log(avg30), xlab = "Elevation", ylab = "log(precip)", 
+      main = "Scatter of log(precip) vs. Elevation")
R> m1 <- lm(log(avg30) ~ ns(CO.elev, df = 3))
R> summary(m1)
R> fits <- m1$fitted.values
R> bad <- is.na(avg30)
R> x <- CO.elev[which(!bad)]
R> ord <- order(x)
R> x <- sort(x)
R> fits <- fits[ord]
R> lines(x, fits, lwd = 3, col = "red")
\end{Sinput}
\end{Schunk}
\begin{Schunk}
\begin{Sinput}
R> resid <- m1$resid
R> hist(resid, freq = F, main = "Histogram of Residuals")
R> lines(density(resid), lwd = 2)
\end{Sinput}
\end{Schunk}

We fit a cubic smoothing spline via least squares to model the relationship between log(precipitation) and elevation. The estimate is shown in Figure~\ref{elev_scatter}, and the histogram of residuals in Figure~\ref{rhist} indicates that a Gaussian assumption is reasonable. We will use the residuals from this model to check for remaining spatial dependence and potential anisotropy. We use \texttt{variog4} to estimate directional sample semivariograms. 

\begin{Schunk}
\begin{Sinput}
R> precip.resid <- cbind(CO.loc[which(!bad), ][, 1], CO.loc[which(!bad), 
+      ][, 2], resid)
R> geodat <- as.geodata(precip.resid)
R> svar4 <- variog4(geodat)
R> plot(svar4, legend = F)
R> legend("bottomright", legend = c(expression(0 * degree), 
+      expression(45 * degree), expression(90 * degree), 
+      expression(135 * degree)), col = 1:4, lty = 1:4)
R> title("Directional Sample Semivariograms")
\end{Sinput}
\end{Schunk}

\begin{figure}
\begin{center}
\includegraphics[width = 3.5 in]{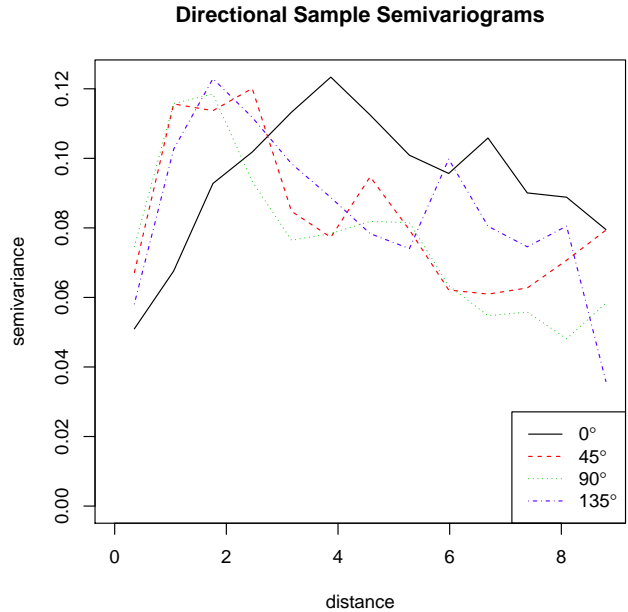}
\end{center}
\caption{A graphical assessment of isotropy in the residuals from the model relating log(precipitation) to elevation.}
\label{raindv}
\end{figure}

The increase, followed by a leveling off, of the semivariance values as distance increases in Figure~\ref{raindv} indicates that there is spatial dependence remaining in the data. We notice that the $0^{\circ}$ semivariogram appears to be slightly different than the other three, but there is no attached measure of uncertainty. We again turn to a nonparametric hypothesis test of isotropy to assist in determining if an assumption of isotropy is reasonable. 

There are two testing procedures for testing isotropy on non-gridded data available in \pkg{spTest}: \citet{guan2004isotropy} and \citet{maity2012test}. To choose between these two, we need to decide whether or not it is reasonable to assume that sampling locations are uniformly distributed on the sampling domain. The methods for non-gridded data from \citet{guan2004isotropy} rely on the assumption that sampling locations are uniformly distributed while the \citet{maity2012test} can be used on any general sampling design. To check this assumption, we can turn to methods from the spatial point process literature to perform a test of complete spatial randomness (CSR) (i.e., a uniform spatial distribution) for the sampling locations. Methods for testing CSR are available in the \R package \pkg{spatstat} \citep{baddeley2004spatstat}. For brevity, we do not include those results here, and we will proceed assuming that CSR holds for these sampling locations.  

For both \citet{guan2004isotropy} and \citet{maity2012test}, we need to choose the lag set, $\bL$, and the contrast matrix, $\A$. The large jumps and decrease in the semivariance values in Figure~\ref{raindv} indicate that semivariogram estimates become unreliable beyond a distance of 2; thus, we should choose lags having Euclidean distance less than this distance. We choose the lag set 
\[
 \bL = \{ \h_1 = (3/4,0), \h_2 = (0,3/4), \h_3 = (3/4,3/4),  \h_4 = (-3/4,3/4)\},
\]
and again we use the matrix $\A$ in Equation~\ref{amat}.
\begin{Schunk}
\begin{Sinput}
R> mylags <- 0.75 * rbind(c(1,0), c(0,1), c(1,1), c(-1,1))
R> myA <- rbind(c(1, -1, 0 , 0), c(0, 0, 1, -1))
\end{Sinput}
\end{Schunk}

The next step to implement the methods from \citet{guan2004isotropy} and \citet{maity2012test} is to determine the size of the moving windows and the block size, respectively, that will be used in estimating the asymptotic variance-covariance matrix, $\mathbf{\Sigma}$. The moving window is shifted over the sampling region, creating subblocks of data used to estimate $\mathbf{\Sigma}$. Likewise, for the test in \citet{maity2012test}, the block size is used to implement the grid-based block bootstrap [GBBB] \citep{lahiri2006resampling}.

There are two steps in making this determination. First, we place a grid over the sampling domain; second, we specify scaling parameters that will define the window/block size in terms of that grid. This necessary two step process for non-gridded locations should be completed keeping three goals in mind: (1) the number of sampling locations per window/block, denoted $n_b$, should be approximately $\sqrt{n}$ \citep{weller2015}; (2) the windows/blocks should have have the same orientation (i.e., square or rectangle) as the entire sampling domain; and (3) the scaling parameters should be compatible with the underlying grid. 

\begin{figure}
\begin{center}
\includegraphics[width = 3.25 in]{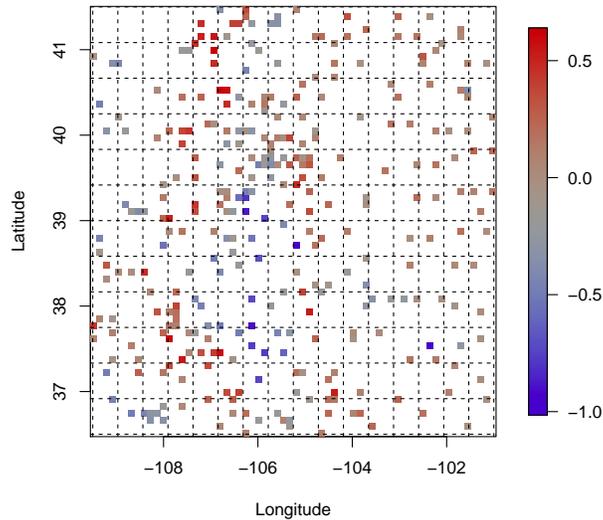}
\end{center}
\caption{Because the sampling locations are not gridded, it can be difficult to assess isotropy properties via a heat map.}
\label{gridplot}
\end{figure}

For the Colorado precipitation data, the dimensions of the sampling region are approximately $8.5^{\circ} \times 5^{\circ}$ (width $\times$ height), providing a total area of 42.5. For $n = 344$ uniformly distributed sampling locations, we expect approximately $344/42.5 = 8.09$ sampling locations per unit area. Recalling goal (1), we seek to construct windows/blocks with $n_b \approx \sqrt{344} = 18.5$, or equivalently, windows/blocks with an area of approximately $18.5/8.09 = 2.29$. Goal (2) indicates we want to create rectangular windows/blocks with slightly larger width than height, and (3) says that if our grid divides the $x$-axis into 12 pieces, then the scaling parameter defining the width of the window/block should be 3 or 4 because those numbers evenly divide 12. For the CO precipitation data, if we choose our grid to divide the $x$-axis into 16 pieces and the $y$-axis into 12 pieces, we have a grid with $(x,y)$ spacing of roughly $(8.5/16, 5/12) \approx (0.53^{\circ}, 0.42^{\circ})$. The resulting grid is plotted in Figure~\ref{gridplot}. Then, choosing our scaling parameters to be $4 \times 3$, we have windows/blocks with dimensions of approximately $(4)(0.53) \times (3)(0.42) = 2.12^{\circ} \times 1.23^{\circ}$ and area of $(2.12)(1.23) \approx 2.61$, or equivalently with $n_b = (2.61)(8.09) \approx 21$.

\begin{Schunk}
\begin{Sinput}
R> quilt.plot(precip.resid[, 1:2], precip.resid[, 3], col = two.colors(n = 256, 
+      start = "blue3", end = "red3", middle = "gray60"), 
+      xlab = "Longitude", ylab = "Latitude")
R> title("Quilt Plot of Residuals and Grid Used for Subsampling")
R> min(precip.resid[, 1])
\end{Sinput}
\begin{Soutput}
[1] -109.483
\end{Soutput}
\begin{Sinput}
R> max(precip.resid[, 1])
\end{Sinput}
\begin{Soutput}
[1] -101.02
\end{Soutput}
\begin{Sinput}
R> min(precip.resid[, 2])
\end{Sinput}
\begin{Soutput}
[1] 36.512
\end{Soutput}
\begin{Sinput}
R> max(precip.resid[, 2])
\end{Sinput}
\begin{Soutput}
[1] 41.467
\end{Soutput}
\begin{Sinput}
R> my.xlims <- c(-109.5, -101)
R> my.ylims <- c(36.5, 41.5)
R> xlen <- my.xlims[2] - my.xlims[1]
R> ylen <- my.ylims[2] - my.ylims[1]
R> my.grid.spacing <- c(xlen/16, ylen/12)
R> xgrid <- seq(my.xlims[1], my.xlims[2], by = my.grid.spacing[1])
R> ygrid <- seq(my.ylims[1], my.ylims[2], by = my.grid.spacing[2])
R> abline(v = xgrid, lty = 2)
R> abline(h = ygrid, lty = 2)
\end{Sinput}
\end{Schunk}

For the functions \texttt{GuanTestUnif} and \texttt{MaityTest}, the upper and lower limits of the sampling region in the $x$ and $y$ directions are given by the \texttt{xlims} and \texttt{ylims} arguments. Note that the values defining the upper and lower limits should be slightly larger than the minimum and maximum observed $x$ and $y$ coordinates. Additionally, the spacing of the grid laid on the sampling region is defined by the \texttt{grid.spacing} argument, and the scaling parameters that define the size of the moving window and blocks in terms of the underlying grid are given by the \texttt{window.dims} and \texttt{block.dims} arguments, respectively. We recommend using visualizations of different grid choices and algebraic calculations, as done above, to assist in choosing a grid and window/block dimensions. When the scaling parameters defining the moving window or block dimensions are not compatible with the number of rows or columns of gridded sampling locations or the dimensions of the grid laid on the sampling region for non-gridded locations, the functions in \pkg{spTest} will print a warning message because they do not currently handle partial (incomplete) blocks. Likewise, if the window or block dimensions are too small for non-gridded sampling locations, the functions \texttt{GuanTestUnif} and \texttt{MaityTest} will discard (sub)blocks that do not have enough sampling locations and print a warning message. The $p$~value of the hypothesis test will be sensitive to the choice of moving window and block dimensions. See \citet{weller2015} and the original works for more recommendations on choosing these values.

The next step for implementing the test in \citet{guan2004isotropy} is choosing the smoothing (bandwidth) parameters for smoothing over lags on the entire domain and within each subblock created by the moving windows. The smoothing parameters should be chosen based on the number and density of the sampling locations with larger values of the smoothing parameter inducing higher levels of smoothing, i.e., averaging over all directions. In our experience, smoothing parameter values between 0.6 and 0.9 tend to produce reasonable results when using a Gaussian smoothing kernel truncated at 1.5. However, the $p$~value of the hypothesis test will change with the bandwidth. For this example, we choose a bandwidth of \texttt{h = 0.7} for smoothing over lags on the entire domain and on the subblocks (\texttt{sb.h}) of data created by the moving window. We also use the default Gaussian smoothing kernel (\texttt{kernel = "norm"}) truncated at 1.5 (\texttt{truncation = 1.5}).

Finally, for the test in \citet{maity2012test} we need to choose the number of bootstrap resamples that will be used in the GBBB procedure to estimate $\mathbf{\Sigma}$. We recommend using at least 50 bootstrap samples; however, the bootstrapping procedure is computationally intensive. We choose \texttt{nBoot = 100} bootstrap samples for our example, and we note that the number of bootstraps does not affect the precision of the $p$~value, which is computed via the asymptotic $\chi^2$ distribution. Having determined values for the different options, we can now perform the hypothesis tests.

\begin{Schunk}
\begin{Sinput}
R> myh <- 0.7
R> myh.sb <- 0.7
R> tr.guan <- GuanTestUnif(spdata = precip.resid, lagmat = mylags, 
+      A = myA, df = 2, h = myh, kernel = "norm", truncation = 1.5, 
+      xlims = my.xlims, ylims = my.ylims, grid.spacing = my.grid.spacing, 
+      window.dims = c(4, 3), subblock.h = myh.sb)
R> tr.guan$p.value
\end{Sinput}
\begin{Soutput}
p.value.chisq 
    0.3608579 
\end{Soutput}
\begin{Sinput}
R> tr.maity <- MaityTest(spdata = precip.resid, lagmat = mylags, 
+      A = myA, df = 2, xlims = my.xlims, ylims = my.ylims, 
+      grid.spacing = my.grid.spacing, block.dims = c(4, 
+          3), nBoot = 100)
R> tr.maity$p.value
\end{Sinput}
\begin{Soutput}
p.value.chisq 
    0.2122808 
\end{Soutput}
\end{Schunk}

For both tests, the residuals do not provide evidence in favor of anisotropy. Thus, in developing a spatial model for the residuals, it may be reasonable to assume isotropy. 

\section{Discussion}\label{discussion}
Choosing a covariance function is an important step in modeling spatially-referenced data and a variety of choices for the covariance function are available (e.g., anisotropy, nonstationarity, parametric forms). The \R package \pkg{spTest} implements several nonparametric tests for checking isotropy properties which avoid specifying a parametric form for the covariance function. One concern regarding the methods in \pkg{spTest} is that they tend to have low power when the anisotropy is weak \citep[see, e.g.,][]{weller2015}.

After determining whether or not an assumption of isotropy is reasonable, we can choose a parametric or nonparametric model for the covariance function. \citet{weller2015} include an illustration of the process of determining and modeling isotropy properties and how nonparametric tests of isotropy can be used in this process. We have demonstrated how graphical techniques and the functions in \pkg{spTest} can be used in a complementary role to check for anisotropy. Future work includes extending the functionality of \pkg{spTest} to allow implementing the tests on non-rectangular sampling domains. Additionally, computational efficiency can be improved by programming functions in \proglang{C++}.
\\
\\
\textbf{Acknowledgements}
\\
\\
Weller's work was supported by the National Science Foundation Research Network on Statistics in the Atmospheric and Ocean Sciences (STATMOS) through grant DMS-1106862. We wish to thank the North American Regional Climate Change Assessment Program (NARCCAP) for providing the data used in this paper. NARCCAP is funded by the National Science Foundation (NSF), the U.S. Department of Energy (DoE), the National Oceanic and Atmospheric Administration (NOAA), and the U.S. Environmental Protection Agency Office of Research and Development (EPA).

\newpage

\bibliographystyle{apa}
\bibliography{spTestReferences}

\end{document}